# Analytical prediction of sub–surface thermal history in translucent tissue phantoms during plasmonic photo–thermotherapy (PPTT)

Purbarun Dhar[$, #], Anup Paul[$, ▲], Arunn Narasimhan and Sarit K Das[*]

Department of Mechanical Engineering, Indian Institute of Technology, Madras, Chennai – 600 036, India

[#]E–mail: pdhar1990@gmail.com

[▲]E–mail: catchapu@gmail.com

[*]Corresponding author: E-mail: skdas@iitm.ac.in

[$] These authors have contributed equally to this research work.

## Abstract

Knowledge of thermal history and/or distribution in biological tissues during laser based hyperthermia is essential to achieve necrosis of tumour/carcinoma cells. A semi–analytical model to predict sub–surface thermal distribution in translucent, soft, tissue mimics has been proposed. The model can accurately predict the spatio–temporal temperature variations along depth and the anomalous thermal behaviour in such media, viz. occurrence of sub-surface temperature peaks. Based on optical and thermal properties, the augmented temperature and shift of the peak positions in case of gold nanostructure mediated tissue phantom hyperthermia can be predicted. Employing inverse approach, the absorption coefficient of nano-graphene infused tissue mimics is determined from the peak temperature and found to provide appreciably accurate predictions along depth. Furthermore, a simplistic, dimensionally consistent correlation to theoretically determine the position of the peak in such media is proposed and found to be consistent with experiments and computations. The model shows promise in predicting thermal distribution induced by lasers in tissues and deduction of therapeutic hyperthermia parameters, thereby assisting clinical procedures by providing *a priori* estimates.

**Keywords:** Laser hyperthermia, tissue mimics, bio heat transfer, gold nanostructures, graphene, PPTT, analytical model

# 1. Introduction

Treatment of surface or sub-surface benign tumours or carcinoma cells inducing cellular necrosis by hyperthermia (Ritchie et al., 1994, Simanovskii et al., 2006 and Lapotko et al., 2006) holds promise as a potential clinical therapy and is an area of active research in the biomedical engineering community. The generated heat is localized within the diseased tissue region which ablates the cells and several techniques of hyperthermia therapy have been proposed and researched upon. The more prominent ones are magnetic nanoparticle and field assisted hyperthermia (Hilger et al., 2002, Dombrovsky et al., 2012, Sturesson and Andersson-Engels, 1995 and Jeun et al., 2009), ultrasonic frequency assisted hyperthermia (Floch et al., 1999, Hynynen et al., 1987, Diederich et al., 1999), microwave radiation mediated hyperthermia (Lyons et al., 1984, Billard et al., 1990), laser based photo–thermotherapy assisted by biocompatible, plasmonic nanostructures (O'Neal et al., 2004, Dickerson et al., 2008, Hirsch et al., 2003) etc.. The last method has elicited much attention in recent years due to its prominent non-invasive nature and extensive therapy database generated from animal models. In general, metallic nanostructures exhibiting plasmon response, such as gold and platinum nanostructures, are employed for the same (due to their bio–compatible nature), however, in recent times, graphene has also been reported to exhibit various novel biological effects (Bhattacharya et al. 2014, Dhar et al. 2015) and show promise as a material for enhancing localized thermal ablation/ hyperthermia (Markovic et al., 2011, Yang et al. 2010). However, laser based non–invasive ablation is only applicable to surface or shallow sub–surface tumours/ carcinoma and therefore penetration of the heat pulse and subsequent diffusion requires accurate understanding. Proper insight into the spatio–temporal thermal distribution within tissues during laser ablation therapy is critical to enable necrosis of the diseased cells and ensure safety of healthy neighbouring tissues from thermal damage. Also, since the laser beam traverses deep within the tissue, the thermal transport characteristics along depth requires proper estimation to gauge the extent of damage to cells beneath the surface tumour. Thus, a predictive approach to determine the extent of heat diffusion within the tissue for a therapy event is required to estimate protocol parameters such as ablation time, nanostructure concentration, laser power, etc. is essential.

However, research on the subject matter has been mostly restricted to *in-vitro* experiments and/or full scale computations on tissue phantoms (Ghosh et al., 2014, Paul et al., 2014b, Jaunich et al., 2008) or *in-vivo* animal trials (von Maltzahn et al., 2009, O'Neal et al., 2004, Yang et al., 2010), wherein determining thermal distribution is often non–feasible and the general approach so far has been to record the transient rise in surface temperature, both in the presence and absence of plasmonic nanostructures. Furthermore, knowledge of detailed sub–surface thermal distribution requires extensive 3D simulations which are computationally expensive and time consuming and are of little use to clinicians who would require proper and prompt estimates to optimize an imminent therapy process. The computations become further challenging when plasmonic nanostructures are involved and hence a prompt yet physically consistent approach to determine thermal distribution is required for quick estimates in real–time scenarios. The present article proposes a simplistic semi–analytical methodology to solve the governing heat transport equations to determine the spatio-temporal sub-surface thermal distribution within the tissue phantoms undergoing laser ablation. The regions of sub-surface temperature peak and thermal distribution in case of gold and graphene nanostructure infused phantoms are also deduced based on optical and thermal properties of the nanostructures and the tissue mimics. The predictions have been observed to be accurate within the regions of interest when compared to experiments as well as full scale simulations. The methodology holds promise in providing theoretical understanding of the process and easy parameter estimation, thereby providing prompt and accurate estimates for fine–tuning subsequent therapy procedures.

## 2. Experimental methodologies

The experimental methodology and instrumentation for the present work are similar to those reported in earlier works by the authors' (Sahoo et al., 2014, Ghosh et al., 2014, Paul et al., 2014b) and a schematic is shown in Fig. 1. The tissue phantom utilized is agar gel which is prepared by mixing agar powder (2.6 wt. %) to deionized water heated to ~ 80 °C and continuously stirred. The solution was then heated for 10-15 min at the same temperature with stirring until a gelatinous liquid is formed. In case of nanostructure infused tissue phantom, the nanomaterials are added in the required quantity to the agar gel during the preparation process and further stirrer for 15-20 min to obtain homogeneous distribution of the particles in the agar matrix. The solution is then poured into the experimental test section and kept undisturbed for 2 h for solidification.

A diode–pumped, 1064 nm, near infrared (owing to maximum absorptivity by tissues in this region of the spectrum) laser was used as the source. Near infrared is the region of spectrum for which the specific absorptivity for the tissue material is much higher than that of the blood and hence is a wavelength band of choice for hyperthermia. Thermocouples were placed at different radial locations which were separated by a distance of 2.5 mm along the depth. The above arrangement (as illustrated in Fig. 1) minimizes the error due to closely placed thermocouples. However, an error in predictability of position still persists and intricate details occurring within the depth–wise thermal profile cannot be measured if they occur in the regions between two neighbouring thermocouples. The temperature of the phantom surface at the region where the laser beam is incident has been measured by an infrared camera. Thermocouple has not been employed for the surface for two major reasons, a) during solidification, the shrinkage of the agar often leads to the thermocouple rupturing the top surface and b) the surface thermocouple often blocks major portions of the incident beam and proper heat generation is not observed in sub–surface regions. The process was also computationally simulated in accordance with methodology (Paul et al., 2014a, Ghosh et al., 2014) using commercial solver Comsol 4.2. The computational domain was taken as a cylindrical section of tissue phantom as shown in Fig. 1 and grid independence was performed (Ghosh et al., 2014).

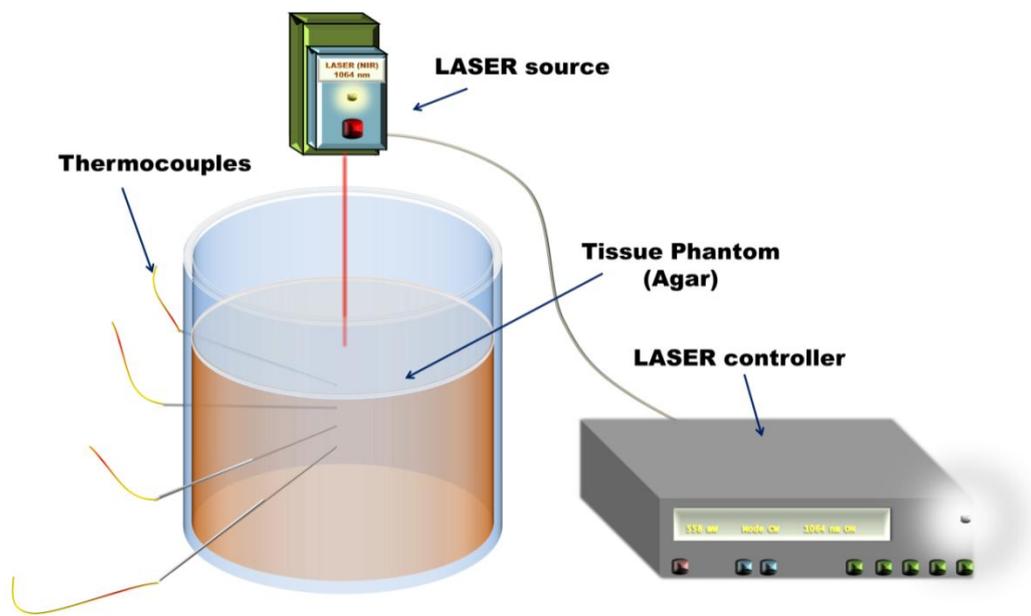

**Figure 1:** The experimental setup. The laser utilized is a diode–pumped solid state 1064 nm near infrared laser. Thermocouples are 1.5 mm J type connected to data acquisition system.

## 3. Theoretical formulation

The mathematical model presented in this article is essentially a simplistic solution to the heat transport equation. The governing thermal transport equation for a conductive system exposed to heating by an incident laser beam is expressible as

$$k\nabla^2 T + Q_{laser} = \rho C_p \frac{\partial T}{\partial t} \qquad (1)$$

where, '$T$', '$Q_{laser}$', '$k$', '$\rho$' and '$C_p$' represent the temperature, laser induced volumetric heat generation, thermal conductivity, density and the specific heat respectively. The heat conduction equation has been solved by disassembling it into two equations, a) a pseudo–steady state equation at the termination of thermotherapy process and b) a transient equation tracing the evolution of temperature at various depths of the tissue at a particular time frame. Since experiments have been performed on tissue phantoms, the metabolic thermo-generation as well as the heat perfusion by blood micro–capillaries has not been considered. The objective lies in obtaining a purely steady state and another purely transient governing equation for the phantom so as to arrive at the two extreme possible solutions for the temperature distribution within the system. The two equations are solved independently to obtain semi-analytical solutions, which are further aggregated back utilizing thermophysical properties based weighted mean to obtain the actual thermal distribution within the tissue phantom during the ablation process. The spatial form of the equation considering attenuation of the beam intensity along depth (Welch, 1984) for the laser induced heat generation is expressed in Eqn. (2). In this case, the tissue has been considered to attain a steady state at the termination of laser thermotherapy and all transient effects are assumed to be non-existent. Basic intensity attenuation model has been used in the spatial component of the equation since after achieving steady state, the heat generation due to multiple scattering (Ghosh et al., 2014) plays no further role in enhancement of temperature. Once the laser is switched off, the system only transmits the absorbed energy within itself and behaves as a purely absorbing system. Furthermore, this essentially reduces the complexity of the spatial transport equation and allows obtaining an analytical solution. The distribution of photon intensity (of a laser beam with Gaussian energy profile) within the tissue mimic at any spatial location can be expressed in accordance to Beer–Lambert law (Welch, 1984) as

$$k\nabla^2 T + (\alpha I_0)\exp\left\{\frac{-r^2}{2\sigma^2 \exp(-\beta z)}\right\}\exp\{-(\alpha+\beta)z\} = 0 \qquad (2)$$

where, '$I_0$', '$\alpha$', '$\beta$' and '$\sigma$' are the beam intensity, the absorption and scattering coefficients of the tissue phantom and the standard deviation governing the decay of intensity within the translucent medium respectively.

The single and multiple scattering models for light–tissue interactions essentially differ with respect to the nature of photon propagation across the translucent medium (Ghosh et al., 2014). In the present context, single scattering has been used analogous with forward scattering, i.e. the photons are scattered mostly along the direction of light propagation or orthogonal to it and this of course diminishes as the intensity attenuates. While the Welch model also incorporates the scattering coefficient, it is dependently coupled to the absorption coefficient and hence the model behaves predominantly as an absorption model. In contrast, in case of the multiple scattering model, the medium also scatters a large portion of the photons against the incoming beam and the incoming and backscattered photons interact leading to higher heat generation. The tissue phantom molecules thereby interact more with the photons and heat generation is higher. In the presence of plasmonic nanostructures, the scattering caliber enhances. Accordingly, the thermal distribution is interplay of the absorption and scattering characteristics behaving as independent components and at the point where both are equally dominant, a temperature peak is observed. A representation of the single and multiple scattering effects with and without nanostructures has been illustrated in Fig. 2.

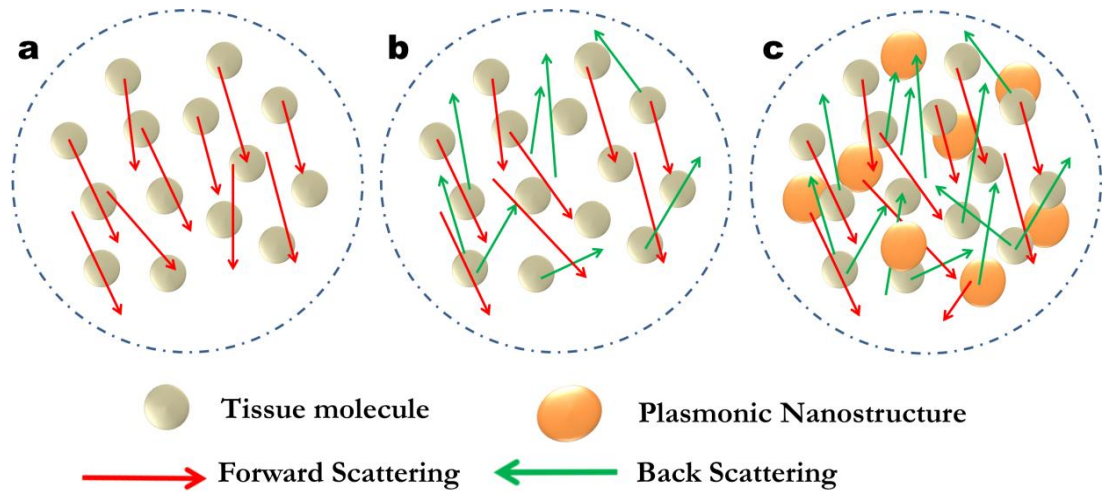

**Figure 2:** Representation of the single scattering (forward scattering) light-matter interactions in **(a)** bare tissue phantom and multiple scattering (forward and backscattering) light-matter interactions in **(b)** bare phantom and **(c)** tissue phantom impregnated with plasmonic nanostructures.

The present approach solely encompasses unidirectional thermal effects in the tissue phantom (along depth, '$z$', considered positive along the direction of laser impingement on the surface) in order to obtain semi–analytical solution for the equation. For the present experiments, the beam diameter at the point of incidence on the tissue phantom surface is 4 mm. A thorough survey of literature shows that in most cases, the beam diameter is very small so as to obtain maximum amount of heat generation over a small region of the tissue due to the higher radiant flux for a smaller beam. It can thus be assumed that such a small beam diameter will lead to considerably greater optical and thermal penetration along the depth as compared to radial diffusion. Thereby, a quasi 1D model can be justified. Moreover, as mentioned, the model is semi-analytical in nature and is aimed at finding prompt estimates of the thermal distribution along depth so as to aid therapy. Fast estimates are difficult to obtain through full scale simulations and analytical solutions cannot be obtained for a 3D system. Hence, a 1D assumption has been resorted to, which the narrow beam width aids. The radial spatial component '$r$', comprising the x and y components, is thereby reduced to zero, and assuming isotropic thermal conductivity for the tissue phantom along depth, the equation reduces to its final form as

$$\frac{d^2T}{dz^2} + \left(\frac{\alpha I_0}{k}\right)\exp\{-(\alpha+\beta)z\} = 0 \tag{3}$$

The second order equation is solved based on two boundary conditions that are expected to exist at the time frame wherein the system has been considered to attain a pseudo-steady state, i.e. at termination of thermo-therapy ($t=t_f$). At this instant, a Dirichlet type temperature boundary condition ($T_{Surf,\ t=t_f}$) is imposed on the surface (subscripted as '$surf$'), as deduced from thermo-therapy experiments. In cases wherein temperature data at any intermediate time frame during heating is required, that particular time frame is considered as $t_f$ to obtain that particular solution. Furthermore, the tissue phantom is modelled as a semi–infinite thermo–sink, such that diffusive transport of heat at very high magnitude of depths is considered to be negligible during the event of thermo-therapy. Thus, a Neumann type adiabatic pseudo-boundary condition is imposed to the phantom, such that as the limits of depth tend to infinity and the thermal gradient along depth decays. Based on the above discussion, the solution for Eqn. 3 can be expressed as

$$T_S(z, t=t_f) = T_{surf, t=t_f} + \left[\frac{\alpha I_0}{k(\alpha+\beta)^2}\right]\{1-\exp(-(\alpha+\beta)z)\} \tag{4}$$

The solution for the steady component, when evaluated at the time corresponding to termination of thermo-therapy with the surface temperature at that time frame as the sole experimental input, yields the spatial pseudo thermo–history within the tissue phantom at that time instant.

The temporal variance of subsurface temperature within the tissue phantom can be evaluated from the transient analysis of the energy equation (Eqn. 1). It is remodelled to mimic a system wherein the thermal wave travels only due to temporal laser induced heating and the progressive depth-wise diffusivity along space is considered to be absent. For such a system, the spatial thermal gradients disappear and the transient equation is expressed as

$$\rho C_p \frac{dT}{dt} - (\alpha I_0) \exp\left\{\frac{-r^2}{2\sigma^2 \exp(-\beta z)}\right\} \exp(-\alpha z)\{1 - \exp(-\beta z)\} = 0 \quad (5)$$

The form of Eqn. 5 implies that it can track the thermo-evolution at a point within the tissue and hence the process of thermal transport is not solely governed by absorption. The tissue, owing to its optical and thermal properties, also participates in the process of scattering the absorbed incident radiation to the surrounding regions and towards the direction of the radiant source itself, leading to a phenomenon termed multiple scattering (Ghosh et al., 2014, Sahoo et al., 2014), which in turn leads to the formation of sub–surface zone of temperature maxima. The laser induced thermo-generation source term in eqn. 5 has been modelled based on the modified optical parameters for the tissue phantom (DasGupta et al., 2009), to incorporate the multiple scattering effect. Similar to the spatial equation, the component of diffused heat in the radial direction is neglected, leading to the final form (assuming isotropic thermal diffusivity and volumetric heat capacity) of the transient equation as

$$\frac{dT}{dt} - \left(\frac{\alpha I_0}{\rho C_p}\right) \exp(-\alpha z)\{1 - \exp(-\beta z)\} = 0 \quad (6)$$

The equation is solved based on the temperature of the tissue phantom just prior to the inception of the thermo-therapy taken as the initial condition ($T_i$). The solution for thermal evolution at any depth, $z_c$, within the tissue phantom for duration of thermo-therapy ($t$) can be expressed as

$$T_T(z=z_c,t) = T_i + \left[\left(\frac{\alpha I_0}{\rho C_p}\right)\exp(-\alpha z)\{1-\exp(-\beta z)\}\right]t \qquad (7)$$

Equations 4 and 7 provide the temperature along depth for a frozen frame of time and the evolution of temperature with time at a particular depth, respectively. When effectively conglomerated, the two can provide the full thermal history within the phantom, for any depth and for any given duration of heating, provided the initial temperature of the phantom and the surface temperature of the phantom at the frozen instant of time are known. However, since the methodology is semi–analytical in nature, the two solutions cannot be averaged. Instead, a thermophysical properties based weighted average ($T(z,t)$), as expressed in Eqn. 8, has been found to consistently yield accurate predictions.

$$T(z,t) = \frac{T_S(z,t=t_f)Bi + T_T(z=z_c,t)Fo}{Bi+Fo} \qquad (8)$$

In Eqn. 8, Bi and Fo represent the Biot and the Fourier numbers associated with the heat transfer process in the tissue phantom respectively. The two non-dimensional parameters are determined based on the characteristic length scale of the phantom (in the present case of a cylindrical domain, the length is one sixth of the cylinder diameter) and the characteristic time, considered as the duration of heating. Predictions obtained from the semi-analytical model for bare tissue phantom for different power levels of the incident beam have been compared against data obtained from full-scale 3D computations and experimental data and are shown in Fig. 3.

## 4. Results and discussions

Thermal history in translucent, soft tissue mimics exhibits counter–intuitive characteristics wherein a peak is formed in the temperature distribution in sub-surface regions rather than that expected at the surface (Ghosh et al., 2014). This occurs due to the multiple-scattering effect by the tissue material, especially due to backscattering of the photons. Since the backscattering is not exclusively treated as an independent parameter in the classical Welch model, maximum temperature is predicted at the surface. The present model is observed to accurately predict the phenomenon. However, the predictability of the model is confined to depths just surpassing the region of peak temperature. As the depth increases beyond the peak zone, the model predicts higher values of temperature compared to experiments. Such anomaly can be explained on the basis of the multiple scattering theory and the simplistic nature of the proposed model.

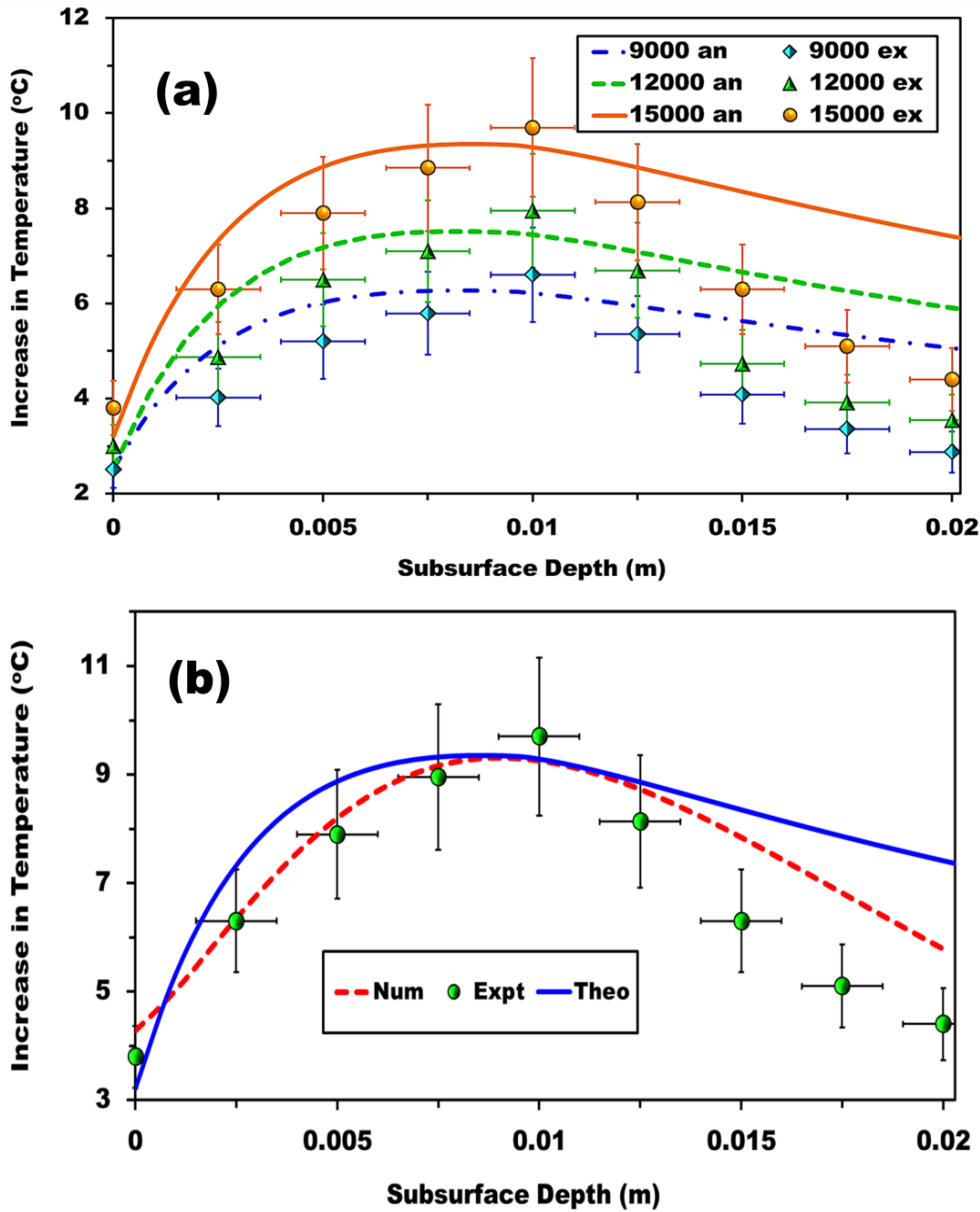

**Figure 3: (a)** Analytical predictions (labelled '*an*') of sub-surface temperature in bare tissue phantoms for various laser power levels compared against experimental data (labelled '*ex*') **(b)** Comparative plot for the analytical solution (labelled '*theo*') with respect to complete 3D simulations ('*num*') against experiments ('*expt*').

The occurrence of temperature peak sub–surface occurs due to the interactivities of the forward and the back scattering. As depth increases below surface, the back scattering effect increases due to increased population of medium molecules that the photons encounter, and consequently the temperature rises. However, with depth, the absorption simultaneously decreases due to intensity attenuation and absorption in the radial directions and thereby the amount of energy available for scattering reduces. At the peak zone, the two effects converge to reach optima, thereby leading to the formation of a temperature peak. Beyond this region, the available energy decays to limits where scattering cannot predominate and the temperature decreases sharply.

The present model is 1D and, in order to obtain analytical solutions, encompasses neither the beam attenuation nor the absorption in directions orthogonal to the laser incidence. Consequently, the model cannot track the decay of beam intensity with depth. The heat generation term due to laser (Eqn. 2) consists of two exponential terms; the first one models the spreading of the beam within the tissue and consequent attenuation and the second term models the absorption and scattering by the tissue. In the present 1D model, the radial component is reduced to zero and hence the exponential term modeling attenuation is reduced to unity. Therefore, the 1D model cannot track attenuation or decay of beam intensity within the tissue. This leads to prediction of higher degrees of heat generation below the critical depth and this is why the model predicts higher temperatures below critical depth. In reality, the numerical model also cannot fully model the attenuation and hence its performance beyond critical depth is also bad.

The intensity thereby remains constant throughout the depth and equivalent to the intensity at the surface. Mathematically, this leads to prediction of enhanced values of temperature in regions wherein the intensity is highly diminished, i.e. at depths beyond the temperature peak, as observed in Fig. 3(a). In the region below the surface and up to the peak zone, the model accurately predicts the thermal history, the accuracy comparable to the predictions from intensive 3D simulations, as shown in Fig. 3(b).

Beyond the peak, the sharp decay of temperature is captured by the simulations, however, only qualitatively. Quantitatively, the computational predictions also fail to predict the thermal response accurately, which is further amplified in case of the analytical predictions. However, as far as predictability of the peak temperature and its location and thermal history up to the peak are concerned, the analytical model shows great potential for accurate mapping of sub–surface thermal history in such media.

During laser aided photo–thermal heat deposition, the penetrability of the photons through the soft matter is a function of the optical response characteristics of the tissue phantom material(DasGupta et al., 2009). Among other optical properties, the absorption and scattering coefficients of the material are of utmost importance for collimated beam based heating. Interestingly, these two properties are tunable and can be tuned to required magnitudes by addition of optically responsive materials; for example, plasmonic nanostructures; to the soft matter. Such nanostructures interact with the bombarding photons and by virtue of surface plasmon response from their surface electron clouds, lead to highly concentrated, localized thermo-generation, which via subsequent absorption and scattering, lead to higher temperatures within the soft matter for the same duration and power of irradiation. Given that the absorption characteristics (~ 40 $m^{-1}$) of NIR radiation (utilized considering *in–vivo* observations(DasGupta et al., 2009), (Ghosh et al., 2014)) for such tissue mimics is nearly an order of magnitude less compared to its scattering potential (~ 530 $m^{-1}$), addition of nanostructures to augment optical and thermal efficiency is a highly sought methodology in the field of Plasmonic Photo–Thermo–Therapy (PPTT)(Schmidt et al., 2008), (von Maltzahn et al., 2009), (Yang et al., 2010).

The addition of plasmonic nanostructures enhances the absorption characteristics of the tissue mimic, leading to higher heat generation. The present study reports use of two different types of such materials; gold mesoflowers and graphene nanoflakes, both exhibiting very high plasmon response in the NIR window. The absorption coefficient for the particle laden tissue increases to ~ 225 ± 25 m$^{-1}$ and ~ 175 ± 25 m$^{-1}$ for gold mesoflowers (Au) concentrations of 0.3 and 0.2 wt. % respectively (Ghosh et al., 2014). However, the scattering coefficient remains more or less equal to that of bare tissue. The enhanced absorption coefficient leads to higher heat generation within the tissue phantom, leading to higher sub–surface temperatures. Furthermore, the enhanced rise in temperature leads to shift of the peak zone towards the surface. The predictability of the model for Au laden tissue mimics against experimental observations and the comparative with computations have been illustrated in Fig. 4(a) and 4(b) respectively.

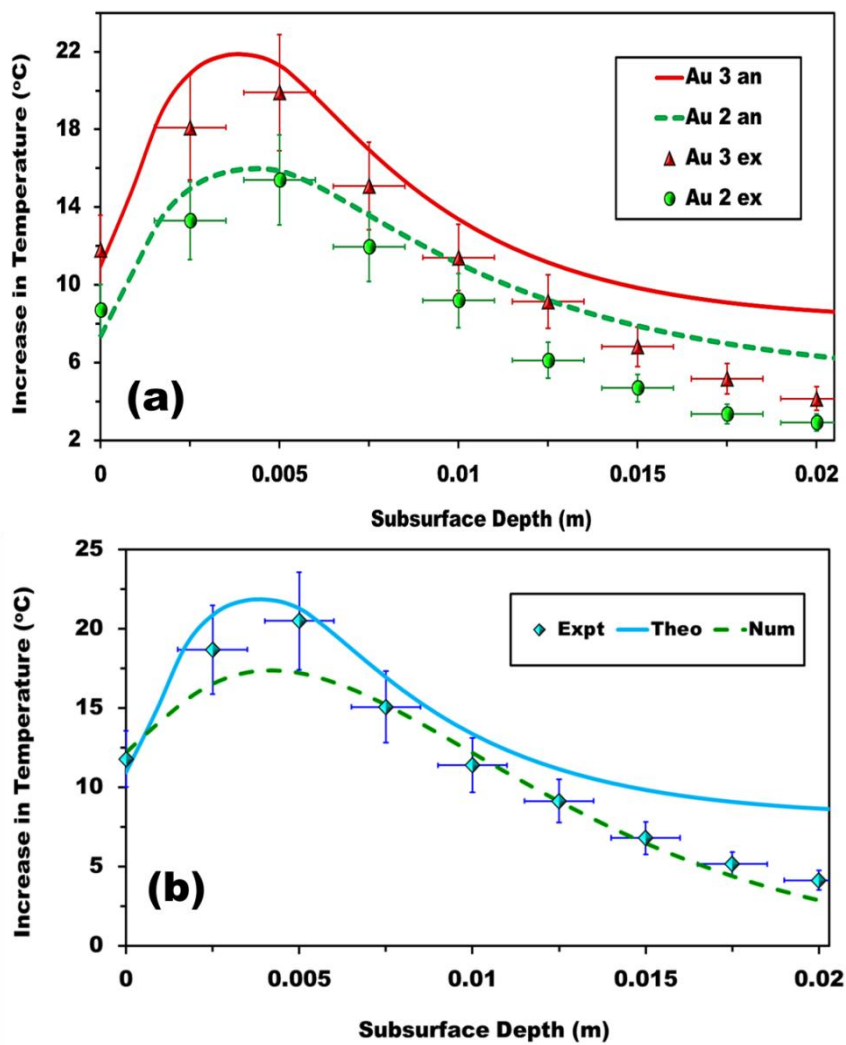

**Figure 4: (a)** Analytical predictions (labelled '*an*') of sub-surface temperature in Au mesoflowers infused bare tissue phantoms for various concentrations (3 mg and 2 mg per gram phantom) compared against experimental data (labelled '*ex*') **(b)** Comparative plot for the analytical solution (labelled '*theo*') with respect to simulations ('*num*') against experiments ('*expt*').

The present study also includes a dimensionally consistent, intuitive correlation to determine the depth at which peak temperature is expected to manifest upon exposure to surface heating, given that '$\alpha$' and '$\beta$' for the translucent soft matter is known. It is possible to deduce the peak depth ($z_p$) theoretically for such substances; the correlation being expressed as

$$z_p \simeq \sqrt{\frac{2}{\alpha \beta}} \qquad (9)$$

The theoretical predictions for peak depth for bare and Au infused tissue phantoms have been compared against experimental and computational results in Fig. 5 and have been found to be accurate.

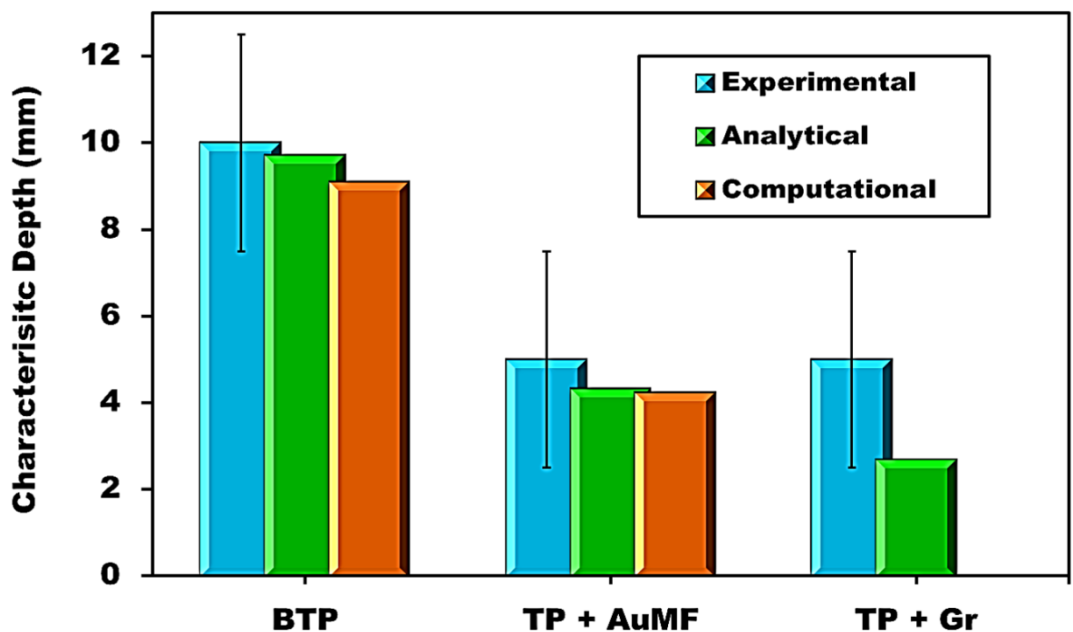

**Figure 5:** Analytical predictions position of sub-surface temperature peak against computational results and experimental data. TP, B, MF and Gr refer to tissue phantom, bare, mesoflowers and graphene respectively. The model also accurately predicts the findings reported (Ghosh et al., 2014) in literature.

Optical and thermal properties of tissue phantoms embedded with graphene nanoflakes are non–existent in literature and an inverse methodology has been adopted to determine the same. As observed from Figures 3(a) and 4(a), the present model can predict the temperatures below surface up to the peak temperature with good accuracy. Furthermore, as expressed in Eqn. 9, it is possible to determine the peak zone from optical and thermal properties. Assuming that similar to Au laden tissues '$\beta$' remains unaffected, from Eqn. 9, it is possible to express '$\alpha$' in terms of '$z$'. This is an assumption that has been resorted to due to absolute lack of data on the scattering coefficients of nanostructures embedded in biomaterials. Scattering coefficient of nanoparticle infused tissues is not reported in literature and determining the same is a complex exercise in itself and beyond the scope of the present article. However, the value of scattering coefficient is expected to be higher than that of bare tissue due to the plasmon resonance by the delocalized pi electron cloud of graphene which would scatter the incoming photons greatly. Hence the present assumption provides estimates of the thermal process for a 'worst case scenario'; something which an important parameter for any estimate.

Utilizing the expression for '$z$' in Eqns. 4 and 7, and obtaining two equations from Eqn. 8 for the experimental peak and immediate neighbouring point temperatures, it is possible to deduce '$\alpha$' for tissue phantoms infused with graphene nanoflakes and the present analysis yields a value of ~ 600 m$^{-1}$ for the same. Based on the obtained value, the peak depth is determined from Eqn. (9) and compared against experiments in Fig. 5. Similarly, the complete sub–surface thermal history is predicted and compared against experiments in Fig. 6, wherein good accuracy is observed. This essentially establishes that the obtained value of 'α' for graphene infused phantoms is accurate. However, the enhanced absorptivity should lead to shift of the peak zone towards surface as compared to Au. This is not observed in experiments, which may be attributed to the spacing of the thermocouples. The predicted peak position however lies within the ranges of positional uncertainty (Fig. 6) and thereby the experimental deviations are justified.

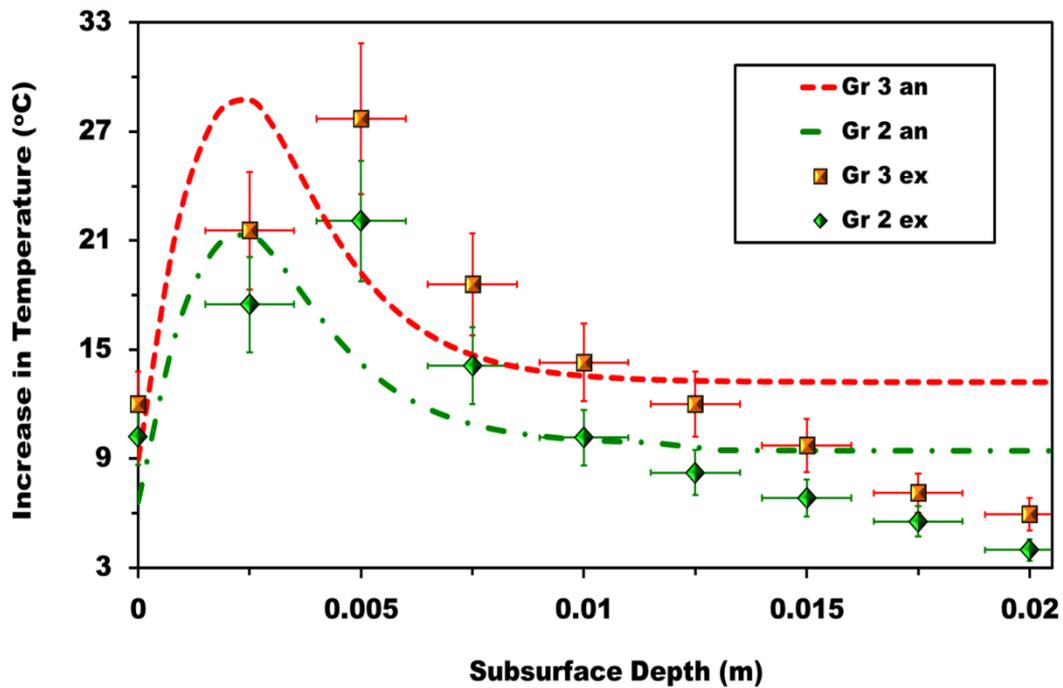

**Figure 6:** Analytical predictions (labelled '*an*') of sub-surface temperature in Graphene infused bare tissue phantoms for various concentrations (3 mg and 2 mg per gram phantom) compared against experimental data (labelled '*ex*').

The deviations observed due to constraints in positioning of the thermocouples will of course lead to some discrepancies for clinicians to estimate the therapy requirements to some extent. However, eliminating such errors is beyond control of the experimental protocol due to limitations posed by the availability and property of materials. In mechanically soft materials like agar, closer spacing of thermocouples (even very fine diameter probes) is very difficult to achieve. Very closed space thermocouples lead to excessive contraction forces within the agar during the solidification process and leads to cracks and tears in the gel around the thermocouples. These have been found to in fact lead to further erroneous results. It is such limitations that researchers resort to full scale simulations which predict the intricate details. However, given the computational requirements, the simplistic model has been proposed and it has been found to mimic numerical results well. As the numerical predictions are also accurate (as observed for most experimental data), it thereby suggests that the analytical predictions also mimic real time data. Hence, the present model is a useful tool for obtaining *a priori* estimates.

## 5. Conclusion

To conclude, a semi–analytical mathematical model to predict the sub–surface thermal behaviour of translucent, soft, tissue mimics has been proposed. The present model is capable of deducing the spatio–temporal variations in temperatures along depth and can also accurately predict the anomalous thermal behaviour in such media; such as formation of sub-surface thermal peaks. Based on known optical and thermal parameters, the model is capable of predicting augmented temperature and shift of the peak positions in case of plasmonic nanostructure based thermo–therapy, such as, gold nanostructure mediated hyperthermia. Based on inverse approach, the absorption coefficient of graphene infused tissue mimics is determined from the peak temperature and found to provide appreciably accurate predictions throughout the depth range. Furthermore, a simplistic yet dimensionally consistent correlation to theoretically determine the position of the peak in such media is proposed and found to be consistent with experimental observations. The present approach shows potential in the prediction of thermal behaviour in tissues and deduction of biological hyperthermia parameters.

Such predictions can reduce the need for extensive experimentations and detailed computations and provide accurate first hand estimates for clinical trials.

## Acknowledgements

The authors thank Dr. Shamit Bakshi of the Department of Mechanical Engineering, Indian Institute of Technology Madras, for the NIR laser. PD also thanks IIT Madras for the post–doctoral fellowship.